# Force-induced Unbinding Dynamics in a Multidimensional Free Energy Landscape

Changbong Hyeon[†]

*School of Computational Sciences, Korea Institute for Advanced Study, Seoul 130-722,*

*Republic of Korea*

**Abstract**

We examined theory for force-induced unbinding on a two-dimensional free energy surface where the internal dynamics of biomolecules is coupled with the rupture process under constant tension $f$. We show that only if the transition state ensemble is narrow and activation barrier is high, the $f$-dependent rupture rate in the 2D potential surface can faithfully be described using an effective 1D energy profile.

[†] Email: hyeoncb@kias.re.kr

Since the birth of chemical dynamics [1, 2], broad classes of simple reactions have been described using a physically suitable one dimensional reaction coordinate [3–5]. It is, however, well appreciated that such a description often fails to capture the dynamics of complex systems such as the folding of proteins or RNA [6]. Interestingly, the response of biological molecules to mechanical force ($f$) is often described using a one dimensional (1D) free energy profile ($F(R)$) with, $R$, the molecular extension that is conjugate to $f$ being the natural reaction coordinate. Use of $R$ as the reaction coordinate is appropriate if the relaxation dynamics associated with all other degrees of freedom are much faster than the dynamics associated with $R$. The celebrated phenomenological Bell model [7] and related microscopic models [8–11], which assume that bond rupture dynamics or forced unfolding of proteins and RNA can be described using $F(R)$, have apparently been adequate in interpreting a number of single molecule experiments. When subject to a tension the transverse fluctuations of the molecule are suppressed, which makes it plausible that the dynamics (forced-unfolding or bond rupture) occurs along an effective 1D free energy profile. A broader validity of the adequacy of $F(R)$ was established in the context of a RNA hairpin dynamics subject to $f$ [12]. By using the calculated free energy profile at $f = f_m$, the force at which the probabilities of being in the folded and unfolded states are equal, it was shown that a Bell-type model can be used to quantitatively predict the dynamics at other $f$ values [12]. It is important to decipher whether energy landscape description based on $R$ alone suffices to describe the force dynamics of biomolecules that, in principle, takes place in a multidimensional surface.

Here, we studied the $f$-dependent unbinding rates, $k_{2D}(f)$, over a barrier on a two dimensional free energy surface $F(x,y)$ in which the reaction coordinate $x$ (describing unfolding or bond rupture) is coupled to an auxiliary variable $y$. The following free energy surface (Fig.1) is considered:

$$F(x,y) = -F^{\ddagger}\left[2\left(\frac{x}{x^{\ddagger}}\right)^3 + 3\left(\frac{x}{x^{\ddagger}}\right)^2\right] - \kappa\left[\left(\frac{x}{x^{\ddagger}}\right) - b\right]y^2 - fx^{\ddagger}\left(\frac{x}{x^{\ddagger}}\right). \qquad (1)$$

When $y=0$, $F(x,y)$ is reduced to the cubic potential that is popularly employed as a microscopic model potential by several group [8, 9, 10]. In $F(x,y)$, a harmonic potential is coupled in the orthogonal direction ($y$). Force along the $x$-direction, tilts the potential surface by $-f \cdot x$. In Eq.1 the $x$-coordinate corresponds to the dynamics of $R$ and the internal degrees of freedom is represented by motions along the $y$ variable. (i) In the absence of tension ($f = 0$), $F(x,y)$ has a local minimum at $x=-x^{\ddagger}$ and barrier top at $x = 0$ along the $y = 0$ axis. The height of potential barrier for the escape dynamics of a quasi-particle, which describes the rupture process, is $F^{\ddagger}$. The parameter $b$ determines the 2D geometry of the transition barrier as well as of the local minimum at $(-x^{\ddagger},0)$. The transition barrier and local minimum become broad when $b$ is small (see Fig.1). However, the condition $b > 0$ should be retained for $F(x,y)$ to have a single saddle point. For $-1 < b \leq 0$, $E(x,y)$ forms two saddle points, and for $b \leq -1$ the local minimum at $x=-x^{\ddagger}$ is not stable. (ii) When $f \neq 0$, $F(x,0)$ has a tension-dependent local minimum at $x_0/x^{\ddagger}= (-1-\varepsilon_f)/2$ and a barrier at $x_b/x^{\ddagger}= (-1+\varepsilon_f)/2$ where $\varepsilon_f \equiv \sqrt{1-f/f_c}$ with $f_c = 3F^{\ddagger}/2x^{\ddagger}$. The barrier height at $f$ is $F^{\ddagger}(f) = F(x_b,0) - F(x_0,0) = F^{\ddagger}\varepsilon_f^3$, which vanishes at $f = f_c$.

To calculate the $f$-dependent escape rate of the quasi-particle from the local minimum of $F(x,y)$ in the intermediate-to-high damping limit, we follow Langer's procedure [13, 14], which extended the Kramers' theory to multidimension. The unfolding (or rupture) rate is

$$k = \frac{\lambda_+}{2\pi}\left[\frac{\det F^{(A)}}{|\det F^{(S)}|}\right]^{1/2} \exp(-\beta F^{\ddagger}(f)) \qquad (2)$$

where total energy $F$ is linearized at the saddle ($S$) and the potential minimum ($A$) using

$$F \approx F^{S,A} + \frac{1}{2}\sum_{i,j} \frac{\partial^2 F^{S,A}}{\partial \eta_i \partial \eta_j}(\eta_i - \eta_i^{S,A})(\eta_j - \eta_j^{S,A}). \quad (3)$$

In the 2D problem associated with Eq.1 the phase space points of the saddle and local minimum are $\{\eta^S\} = (x^S, y^S, p_x^S, p_y^S) = (x_b, 0, 0, 0)$ and $\{\eta^A\} = (x_0, 0, 0, 0)$, respectively. The rate constant $k$ amounts to a flux-over-population expression from the steady state solution of a multidimensional Fokker-Planck equation. The $\lambda_+$ value corresponds to the deterministic growth rate at the saddle point from which the trajectory diverges exponentially along the reaction path. To calculate $\lambda_+$, we use the Hamilton's equations of motion for each variable,

$$\dot{\eta}_i = -\sum_j M_{ij} \frac{\partial F}{\partial \eta_j} \quad (4)$$

and linearize the first derivative of $E$ at $S$ using,

$$\frac{\partial F}{\partial \eta_j} = \sum_k \left(\frac{\partial^2 F}{\partial \eta_j \partial \eta_k}\right)_S \delta\eta_k^S = \sum_k e_{jk} \delta\eta_k^S \quad (5)$$

where $\delta\eta_k^S \equiv \eta_k - \eta_k^S$ with $\eta_k = x, y, p_x, p_y$. Thus, $\{\eta\}$ satisfies the first order matrix equation

$$\dot{\eta}_i = -\sum_{j,k} M_{ij} e_{jk} \delta\eta_k \quad (6)$$

where

$$M = \begin{pmatrix} 0 & 0 & -1 & 0 \\ 0 & 0 & 0 & -1 \\ 1 & 0 & m\gamma_{xx} & 0 \\ 0 & 1 & 0 & m\gamma_{yy} \end{pmatrix} \quad (7)$$

and

$$e = \begin{pmatrix} -\dfrac{6F^{\ddagger}}{(x^{\ddagger})^2} & 0 & 0 & 0 \\ 0 & \kappa(2b+1-\varepsilon_f) & 0 & 0 \\ 0 & 0 & 1/m & 0 \\ 0 & 0 & 0 & 1/m \end{pmatrix} \qquad (8)$$

Among the four eigenvalues of Eq.6, the expression for the physically relevant one $\lambda_+$ is

$$\lambda_+(f) = -\dfrac{\gamma_{xx}}{2} + \sqrt{\left(\dfrac{\gamma_{xx}}{2}\right)^2 + \dfrac{6F^{\ddagger}}{m(x^{\ddagger})^2}\varepsilon_f} \qquad (9)$$

The determinants of the Hessian matrices at minimum ($A$;+) and barrier top ($S$; −) are calculated using

$$F^{(\pm)} = \begin{pmatrix} \pm\dfrac{6F^{\ddagger}}{(x^{\ddagger})^2}\varepsilon_f & 0 & 0 & 0 \\ 0 & \kappa(2b+1\pm\varepsilon_f) & 0 & 0 \\ 0 & 0 & 1/m & 0 \\ 0 & 0 & 0 & 1/m \end{pmatrix}, \qquad (10)$$

where $F^{(+)} \equiv F^{(A)}$ and $F^{(-)} \equiv F^{(S)}$. Finally, the escape rate over the 2D model potential can be written as

$$k_{2D}(f) = \dfrac{\lambda_+}{2\pi}\sqrt{\dfrac{b+(1+\varepsilon_f)/2}{b+(1-\varepsilon_f)/2}}\exp\left(-\varepsilon_f^3 \beta F^{\ddagger}\right). \qquad (11)$$

The stringent condition, $b > 0$, ensures that the potential has only a single saddle point. The parameter $\kappa$ in Eq.1, which defines the strength of the harmonic potential in $y$-direction, does not affect the barrier crossing kinetics in $k_{2D}(f)$ because of the symmetry of the cubic potential around the inflection point at $x = -x^{\ddagger}/2$. The rate at zero force $k_0(\equiv k_{2D}(0))$ depends on $b$ as

$$k_0(b) = \dfrac{\lambda_+(0)}{2\pi}\sqrt{\dfrac{b+1}{b}}\exp\left(-\beta F^{\ddagger}\right) \qquad (12)$$

In the high damping limit ($\gamma_{xx} \gg 1$), the above expression is simplified to

$$k_{2D}(f) = \frac{\varepsilon_f}{2\pi m \gamma_{xx}} \frac{6F^{\ddagger}}{(x^{\ddagger})^2} \sqrt{\frac{b+(1+\varepsilon_f)/2}{b+(1-\varepsilon_f)/2}} \exp\left(-\varepsilon_f^3 \beta F^{\ddagger}\right)$$

$$= D_{2D}(b,f) k_0(b) \varepsilon_f \, e^{(1-\varepsilon_f^3)\beta F^{\ddagger}}$$

$$= D_{2D}(b,f) k_{1D}(f) \tag{13}$$

where $k_0(b)$ is the rate at $f = 0$ and $D_{2D}(b,f) = \sqrt{\left(\frac{b+(1+\varepsilon_f)/2}{b+(1-\varepsilon_f)/2}\right)\left(\frac{b}{b+1}\right)}$ is the correction factor for 2D reaction surface. $k_{1D}(f)$ is the rate expression for the 1D microscopic model ($\kappa = 0$ in Eq.1) with $\nu = 2/3$ [8, 9],

$$k_{1D}(f)/k_0(b) = \left(1 - \frac{\nu f x^{\ddagger}}{F^{\ddagger}}\right)^{1/\nu - 1} \exp\left(\beta \Delta G^{\ddagger} \times \left[1 - (1 - \nu f x^{\ddagger}/F^{\ddagger})^{1/\nu}\right]\right). \tag{14}$$

A few comments involving Eq.13 are in place: (i) $D_{2D}(b,f) \approx 1$ is obtained either when $b \gg 1$ or when $\varepsilon_f \approx 1$. The deviation of $D_{2D}(b,f)$ from the unity becomes significant especially as $b$ becomes smaller and $f$ approaches the critical value ($f \to f_c$) so that the free energy barrier is about to vanish. (ii) The parameter $b$ describes the geometry around the saddle point and the local basin of attraction. When $b \gg 1$ the saddle point is sharply defined and transition state ensemble (TSE) is narrow. However, when $0 < b \ll 1$, both TSE and the local basin corresponding to the bound state are broad, leading to a large fluctuations orthogonal to the $x$-coordinate (Fig.1). Description of force kinetics using $k_{1D}(f)$ fails when $0 < b \ll 1$ or the distinction between the transition state and the native basin (or bound state) is not transparent ($f/f_c \to 1$). In both scenarios the actual reaction paths deviate significantly from that determined along a predefined reaction coordinate. Under these conditions the one-dimensional reaction coordinate projected from multidimensional space cannot adequately describe the true dynamics even in the presence of $f$, which can usually moderate such fluctuations.

Instead of using the minimum path of the 2D surface as a 1D reaction coordinate, one can also consider the projection of 2D free energy surface by integrating the fluctuations in the y-coordinate, which allows us to define an effective 1D energy profile,

$$F^{eff}(x;b) = -\beta^{-1}\log\int_{-\delta}^{\delta} dy e^{-\beta F(x,y)}$$

$$\approx F(x,0) + \frac{1}{2}\beta^{-1}\log\left[\frac{\kappa\beta}{\pi}(b - x/x^{\ddagger})\right]. \qquad (15)$$

As shown in Fig.2, $F^{eff}(x;b)$ and $F(x;0)$ differs qualitatively when $b \to 0$ but only differs by a constant ($F^{eff}(x,b) - F(x,0) \approx \frac{1}{2}\beta^{-1}\log\left[\frac{\kappa\beta b}{\pi}\right]$) when $b > 1$. For $F^{eff}(x;b)$ with $b \to 0$ the effective transition barrier is smaller than in $F(x,0)$ (Fig.2), which suggests that in the broad TSE the free energy barrier is lowered when the higher-dimensional free energy surface is projected onto one dimension. As a result the rate would increases provided that the prefactor for 1D- Kramers equation is nearly independent of $b$. In performing the integration to obtain the result in Eq.15 we set $\delta = \infty$. This approximation is only valid for harmonic potential with a large $\kappa$ or large $b$ that results in rapid relaxation. For small $\kappa$ or small $b$, motions along y-axis are slow and hence the $\delta$ value should be finite. Therefore, the barrier $F^{eff}(x,b=0.1)$ in Fig.2 may be slightly underestimated. However, the exact calculation of $D_{2D}(b,f)$ for $b \to 0$ leads to a pathological result, in which $D_{2D}(b,f)$ decreases with $f$. If the behavior of $D_{2D}(b,f)$ at small $b$ is combined with the rest of the term in Eq.(13), $k_{2D}(f)$ exhibits nonmonotonic dependence on $f$. It is of particular note that even in multidimensional version of Kramers rate expression suggested by Langer, the transition path should be well defined along the multidimensional surface; projecting the 2D surface onto 1D profile leads to a physically incorrect result especially for $b \to 0$ in which the flat free energy barrier produces no dominant transition path.

As another plausible scenario where the effect of multidimensionality is manifested in the context of force-induced unfolding kinetics, one can study hydrodynamic interaction that dynamically couples the motions along "*x*"- and "*y*"-coordinates. In the presence of hydrodynamic interactions, the mobility tensor $M$ (Eq.7) is modified into $M_{HI}$ with $\gamma_{xy} \neq 0$,

$$M_{HI} = \begin{pmatrix} 0 & 0 & -1 & 0 \\ 0 & 0 & 0 & -1 \\ 1 & 0 & m\gamma_{xx} & m\gamma_{xy} \\ 0 & 1 & m\gamma_{xy} & m\gamma_{yy} \end{pmatrix} \quad (16)$$

which alters the deterministic growth rate $\lambda_+$ into $\lambda_+^{HI}$, leaving other terms in $k_{2D}(f)$ (see Eq.11) unchanged. Since the analytical expression for $\lambda_+^{HI}$ is quite involved, we obtain $\lambda_+^{HI}$ numerically by varying $\gamma_{xy}$ and $f/f_c$ and plot $\lambda_+^{HI}(\gamma_{xy}, f/f_c)$ in Fig.3. The effect of hydrodynamic interaction on the kinetics through $\lambda_+^{HI}$ is the most significant when the activation barrier ($F^\ddagger$) is small. More significantly, pronounced is the variation of $\lambda_{HI}$ when $f/f_c$ is small (see Fig.3-A). It is of note that hydrodynamic interactions ($\gamma_{xy} \neq 0$) increases the rate of deterministic divergence from the saddle point ($\lambda_+^{HI} > \lambda_+$), which partially compensates the reduced $D_{2D}(b,f)$ due to large fluctuations.

In order to extract meaningful parameters using 1D profiles, the ensemble of reaction paths should go through a deep and narrow "trough" in the multi-dimensional energy landscape (see $b = 10$ case in Fig.1), so that fluctuations due to coupling to the auxiliary coordinates is minimal and that the transition path is well defined. Unless this condition is met, the force-induced rupture kinetics in a multidimensional energy landscape can be drastically different from that inferred from a pre-selected 1D reaction coordinate.

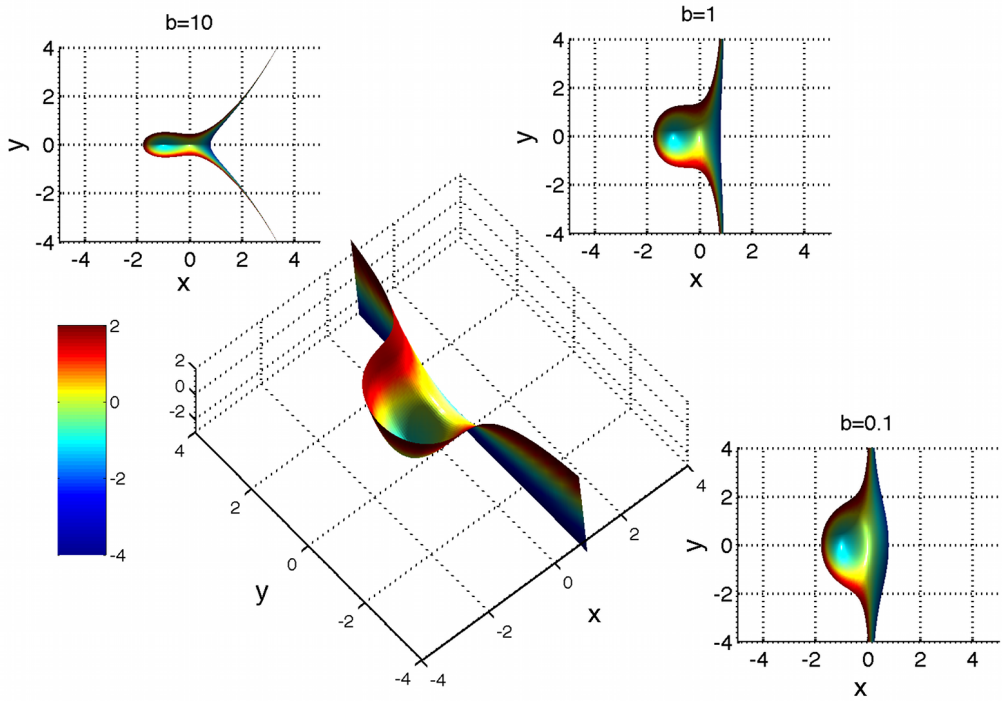

**Figure 1:** A two-dimensional energy surface using Eq.1 with f=0, κ=1 and $\beta F^{\ddagger} = 1$ and varying b>0 values. x and y coordinates are scaled by $x^{\ddagger}$. The energy scale is color-coded in $k_B T$ unit.

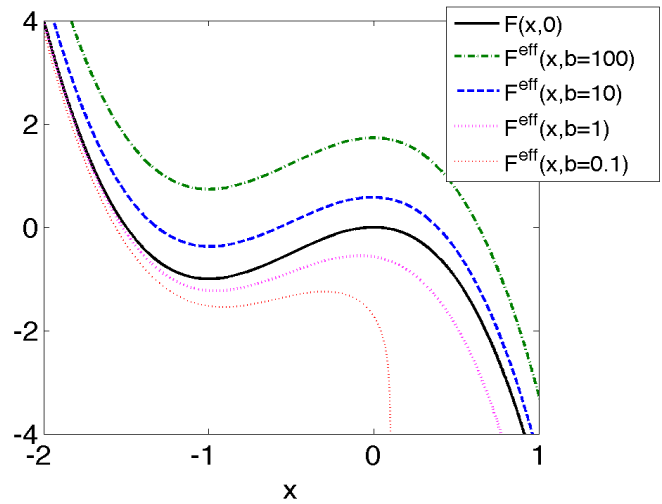

**Figure 2:** The effective 1D free energy profiles projected from 2D surface for varying *b* values. F$^{eff}$(x,b) deviates from F(x,0) when 0<b<1.

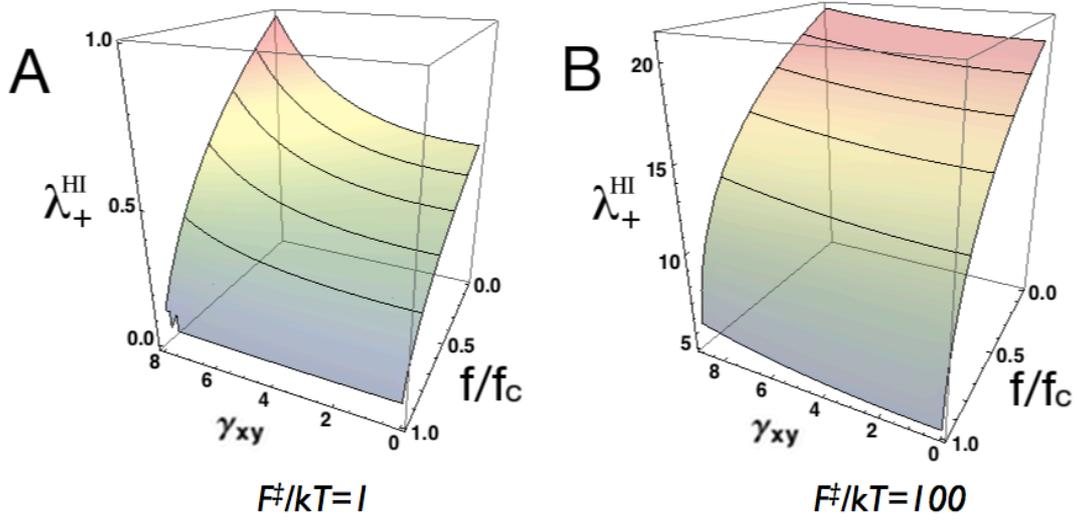

**Figure 3:** The effect of hydrodynamic interaction ($\gamma_{xy}$) on the deterministic growth rate calculated for A. $\beta F^{\ddagger} = 1$ and $\beta F^{\ddagger} = 100$ with other parameters being $\kappa = 1$, $b = 1$, $\gamma_{xx} = \gamma_{yy} = 10$, $x^{\ddagger} = 1$ and $m = 1$.